\documentclass[aps,prf,floatfix,longbibliography]{revtex4-1}
\usepackage[dvipsnames,rgb,dvips]{xcolor}
\usepackage{graphicx}
\usepackage{float}
\usepackage{overpic}
\usepackage{bbm}
\usepackage{amsmath,amssymb}
\usepackage{amsfonts}
\makeatletter
\def\amsbb{\use@mathgroup \M@U \symAMSb}
\makeatother
\usepackage[bbgreekl]{mathbbol}
\usepackage{units,mathrsfs}

\usepackage{array}
\usepackage[color=green!40]{todonotes}
\usepackage[normalem]{ulem}
\usepackage{textcomp}
\usepackage{gensymb} 
\newcolumntype{C}[1]{>{\centering\let\newline\\\arraybackslash\hspace{0pt}}p{#1}}

\newcommand{\beq}{\begin{equation}}
\newcommand{\eeq}{\end{equation}}

 \newcommand{\bfi}{\begin{figure}[h]}
 \newcommand{\efi}{\end{figure}}

\begin{document}


\title{Probability of noise-induced separatrix crossing for inertial particles in flows.}

\author{Jean-R\'{e}gis Angilella}
\address{Universit\'e de Caen Normandie, ESIX, ABTE EA4651-ToxEMAC, 50100 Cherbourg, France}

\vskip.5cm
\begin{abstract}
{ 
The motion of weakly inertial Brownian particles, transported by steady two-dimensional 
fluid flows, is investigated by means of asymptotic  methods. We focus on the phenomenon of noise-induced separatrix crossing, which can force particles to enter or exit recirculation cells in an unpredictable manner.
An analytical expression for the probability of separatrix crossing is obtained. It can be applied to a wide variety of flows, provided some elementary kinematical quantities of the fluid flow are known. It does not require to solve particle trajectories.
}
\end{abstract}
 
\maketitle
  






%
 
\section{Introduction}

Recirculation cells, characterized by closed streamlines embedded in an open flow, are ubiquitous in fluid flows and play an important role in the Lagrangian dynamics. Any fluid point approaching the dividing streamline between both zones has an uncertain dynamics, like a simple pendulum released in the vicinity of its upper equilibrium position. Indeed, small perturbations can cause the fluid point to cross the dividing streamline (also called "separatrix") and leave the open zone to join the closed zone, or vice-versa. This is why recirculation cells are, in practice, separated from the rest of the flow by a stochastic zone rather than a well-defined dividing streamline, even in quasi-steady flows. These uncertainties are observed also for any finite-size particle that would be carried by the flow, like aerosols entrained in the wake of a cylinder \cite{Richmond2006},   micron-sized particles in the context of microfluidic \cite{Haddadi2014}, Brownian particles settling in the vicinity of Stommel cells \cite{Stommel1949,Simon1991}, or pollutants in the wake of large obstacles \cite{Tominaga2016} where turbulence is often treated as a random force, to quote but a few examples.
In general, to predict the proportion of particles which cross the separatrix, one has to perform a large number of simulations involving a large number of particles, to obtain reliable statistical data. This approach is very time consuming, and reveals to be unrealistic when the numerical model does not capture accurately the small-scale features of perturbations. The objective of the present work is to derive an analytical expression for the probability of separatrix crossing, which could be applied to a wide variety of particle-laden flows, without requiring to solve particle trajectories.
 
 Such a general criterion for separatrix crossing has been proposed for non-Brownian aerosols transported in a quasi-steady fluid flow submitted to weak periodic perturbations \cite{Verjus2016}. It can be applied to any two-dimensional flow, provided a single trajectory along the limiting streamline of the steady component of the flow is known either analytically or numerically. 
Nevertheless, this purely deterministic approach fails to predict efficiently the behavior of suspended particles in many real situations where some external weak noise is present. 
The purpose of this paper is therefore to generalize the deterministic result of Ref. \cite{Verjus2016} to the case of a particle moving in a steady flow and submitted to the action of a small  stochastic force.

Since pioneering works done in the eighties \cite{Friedlin1984,Graham1984a,Graham1984b,Kautz1988}, noise-induced escape from attractors and, more generally, the various effects of noise on dynamical systems, have been investigated in many contexts  \citep{Beale1989,Grassberger1989,Bulsara1990,frey1992,Frey1993,
Simiu1996,Simiu1996Melnikov,Landa1996Pendulum,Franaszek1998,soskin2001,Liu2004,Zhu2005,Ibrahim2006}.  In the present work
  attractors are not under study, but we will make use of some of the mathematical techniques developed by these authors to study the effect of noise on the particle dynamics.  
\begin{figure}[h] 
\center{\includegraphics[scale=0.6]{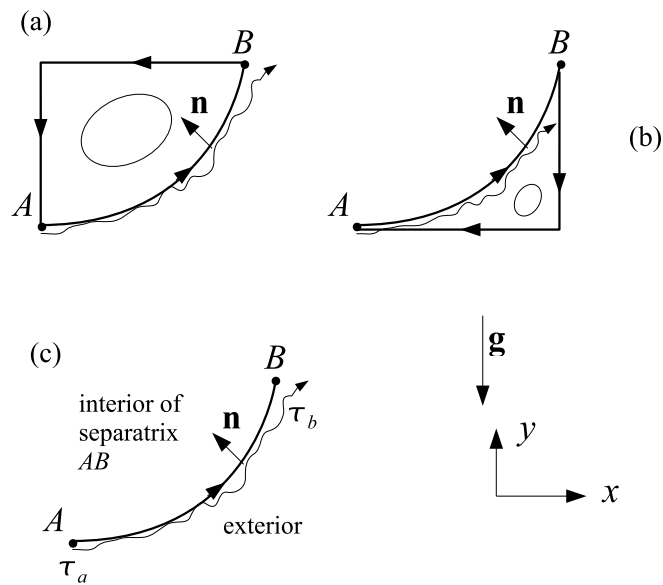}}
\caption{Sketch of a dividing streamline of the base flow $\mathbf{u}_0$ (thick line $AB$). (a): convex recirculation cell ; (b): concave cell. Thick lines are dividing streamlines of the fluid flow, thin lines are particle trajectories.  Vector  $\mathbf n$ is the unit vector perpendicular to the streamline, such that $(\mathbf{u}_0, \mathbf n, \hat{\mathbf z})$ is right-handed.}
\label{SepAB}
\end{figure}

\section{General considerations}

 We consider tiny isolated spherical particles of radius $a$ and mass $m_p$, driven by a fluid with kinematic viscosity $\nu$, and moving in the vicinity of a recirculation cell in a vertical $(x,y)$ plane. 
Even though most recirculation cells are unsteady, we assume that the characteristic time of the flow unsteadiness is large compared to convective times near the cell.  
 The fluid velocity field, denoted as $\mathbf{u}_0$,  is therefore taken to be steady.
The "boundary" of recirculation cells is defined as a dividing streamline $AB$, or set of streamlines (as sketched in Fig. \ref{SepAB}), separating the cell from the external open flow, or from another cell, or a wall. 
Since the fluid flow is  steady, these streamlines are time-independent and the boundary is well-defined.

 We assume that the Reynolds number of the particles remains smaller than unity.
  To describe their motion, it is convenient to define the interior and the exterior of the separatrix according to its curvature (see Fig.\ \ref{SepAB}(c)). Note that the interior and exterior of the separatrix do not always coincide with the interior and exterior of the cell (like in the concave cell of Fig.\ \ref{SepAB}(b)).  
 We use a typical fluid velocity $V_0$, length scale $L_0$, and the particle mass $m_p$ to non-dimensionalize the motion equations. 
 The resulting non-dimensional motion equation of particles much heavier than the fluid, at low particle Reynolds number, reads
 \begin{equation}
 \ddot{\mathbf{x}} =  \frac{1}{Fr}  \hat{\mathbf g}  +
\frac{1}{St}\left(\mathbf{u}_0(\mathbf{x}) - \dot {\mathbf{x}}\right)   + f_0 \boldsymbol{\xi},
\label{ptclDynAdim}
\end{equation}
where 
$
St = \frac{2\gamma_p}{9} \frac{a^2}{\nu} \frac{V_0}{L_0}
$
is the Stokes number, $\gamma_p \gg 1$ is the particle-to-fluid density ratio, $\hat{\mathbf g}$ is the unit vector in the direction of gravity,
 and $Fr = V_0^2/(g L_0)$ is the Froude number. The non-dimensional intensity of the random force $f_0$ is related to its dimensional counterpart $F_0$ by
$
f_0 = {F_0} L_0 /({m_p} {V_0^2}).
$
Vector $\boldsymbol{\xi}(t) =(\xi_1,\xi_2)$ has random components, both of which have the same probability density and unit variance. 
We also introduce the non-dimensional settling velocity of particles in still fluid 
$\mathbf{V}_T = V_T \hat{\mathbf g}$, with
$V_T  =  {St}/{Fr}$.
Throughout the paper, we assume that $Fr$ is larger than or of order unity, and that $St \ll 1$. These assumptions are very common for tiny solid or liquid objects in gases, and manifest the fact that sedimentation is slow and inertia is weak.   Also, we assume that the non-dimensional noise intensity $f_0$ satisfies  
$f_0 \, St \ll 1$, and can therefore reach moderate values.

 Each random component $\xi_i(t)$ in Eq. (\ref{ptclDynAdim}) is taken to be piecewise constant in time, i.e. it is constant and equal to $\xi_{ik}$ over an interval $[\tau_k,\tau_{k+1}]$, where $\tau_k$ is an increasing series of real numbers.  
Whatever the cause of the random force, we assume in this note that its characteristic time is much smaller than convective times, i.e. $\delta\tau_k \equiv \tau_{k+1}-\tau_k \ll 1$. The durations $\delta\tau_k$ will be assumed to be either constant ($\delta\tau_k = \delta\tau$ for all $k$), or random and memoryless, i.e. with an exponential distribution with mean $\delta\tau$. In the latter case, noise corresponds to a Kubo-Anderson process \cite{Kubo1954,Anderson1954,Brissaud1974}. In both cases the noise is of colored type, with a non-zero characteristic time scale $\delta\tau$. The autocorrelation function $\langle \xi(t) \xi(t') \rangle$ 
 is equal to $\exp(-|t-t'|/\delta\tau)$ for exponentially distributed durations. The diffusion coefficient, obtained from the Langevin Eq.\ (\ref{ptclDynAdim}) with $\bf u_0 \equiv \mathbf 0$ and $1/Fr = 0$, satisfies
 $
 D = \alpha \, f_0^2 \, St^2 \, \delta\tau ,
 $
where $\alpha=1/2$ for constant durations $\delta\tau_k$, and $\alpha=1$ for exponentially distributed durations. 

\subsection{Calculation of the undisturbed streamfunction jump}

To analyse the dynamics near separatrix $AB$, we consider the undisturbed non-dimensional streamfunction $\psi_0(x,y)$  defined as  $\mathbf{u}_0 = \nabla \psi_0 \times \mathbf{\hat z}$, where $\mathbf{\hat z}$ is the unit vector in the $z$ direction, perpendicular to the flow plane $(x,y)$. In addition, we assume without loss of generality that the flow goes from $A$ to $B$. The value of $\psi_0$ at any point $\mathbf x$ indicates whether $\mathbf x$ is on the left-hand-side of the dividing streamline $AB$ (i.e. $\psi_0(\mathbf x) > \psi_0(A)$) or on the right-hand-side of $AB$ ($\psi_0(\mathbf x) < \psi_0(A)$).  The calculations presented here are similar to the calculations done for the construction of separatrix maps in homoclinic or heteroclinic cycles \cite{Kuznetsov1997}. More details on the calculations can be found in Ref. \cite{Angilella2019}.
We consider a particle released near point $A$ at time $\tau_a$ and passing nearest to point $B$ at time $\tau_b$. The time-dependent position of this particle is denoted as $\mathbf x(t)$, and obeys Eq.\ (\ref{ptclDynAdim}). We calculate the variation of $\psi_0(\mathbf x(t))$ between $\tau_a$ and $\tau_b$ as follows. By noticing that $d\psi_0(\mathbf x(t))/dt = \nabla \psi_0 \cdot \dot{ \mathbf x}$, and making use of Eq.\ (\ref{ptclDynAdim}), we get:
\begin{eqnarray}
\Delta \psi_0 &\equiv& \psi_0(\mathbf x(\tau_b)) - \psi_0(\mathbf x(\tau_a)) ,\\
&=& \int_{\tau_a}^{\tau_b} \!\!\!\! \nabla \psi_0(\mathbf x(t)) . \left[    - St \, \ddot{\mathbf{x}} + \mathbf {V}_T + St \, f_0 \boldsymbol{\xi}(t) \right] dt.
\end{eqnarray}
Assuming the particle travels very close to the separatrix, we set \cite{GH83,Kuznetsov1997}:
$
\mathbf x(t) \simeq \mathbf q(t-t_0),
$
where $\mathbf q$ is the position of a fluid point traveling on the separatrix, that is: $\mathbf {\dot q} = \mathbf u_0(\mathbf q)$, and $t_0$ is the time where $\mathbf x$ passes nearest to $\mathbf q(0)$ which is chosen arbitrarily on the arc $]A,B[$. Under these hypotheses, we obtain:
\begin{equation}
\Delta \psi_0   
 = \Delta \psi_0^0  
 + St \, f_0  \sum_k     {\delta s}_k \,  {\xi}_{kn},
 \label{Deltapsi01}
\end{equation}
where 
$\delta s_k = |\mathbf{q}(\tau_{k+1}-t_0)-\mathbf{q}(\tau_{k}-t_0)|$ are discrete length elements on the arc $AB$ and $\xi_{kn} =
  {\boldsymbol \xi}_k .  \mathbf{n}$.
The term $\Delta \psi_0^0$ is the  streamfunction variation in the deterministic dynamics, already discussed in Ref. \cite{Verjus2016}:
 \begin{equation}
\Delta \psi_0^0 =      - St \, \left(  \overline{u^2_0}
+  \frac{ x_{AB} }{Fr} \right).
 \label{Deltapsi00}
\end{equation}
The term $\overline{u^2_0}$
  is the curvature-weighted integral of the squared fluid velocity:
  \begin{equation}
 \overline{u^2_0} = {\cal L}_{AB} \, \langle  R^{-1} |\mathbf{u}_0|^2 \rangle_{AB}  ,
  \end{equation}
  where $x_{AB}=x_B-x_A$, $R^{-1} = \hat{\mathbf z} \cdot \dot{\mathbf q} \times
\ddot{\mathbf q} / |\mathbf{u}_0|^3$ is the local curvature of  arc $AB$, and $\langle \cdot \rangle_{AB}$ is the average along $AB$:
  $
\langle X \rangle_{AB} =   (\int_A^B\, X(s)\, ds) / {{\cal L}_{AB}}$, where ${\cal L}_{AB}$ is the total arc-length of $AB$.
  Note that $\overline{u^2_0}$ is  not always positive: it is positive if separatrix $AB$ is anti-clockwise, and negative if it is clockwise. The discrete sum in Eq.\ (\ref{Deltapsi01}) is the contribution of the random force.
  
\begin{figure}[h] 
\center{\includegraphics[scale=0.8]{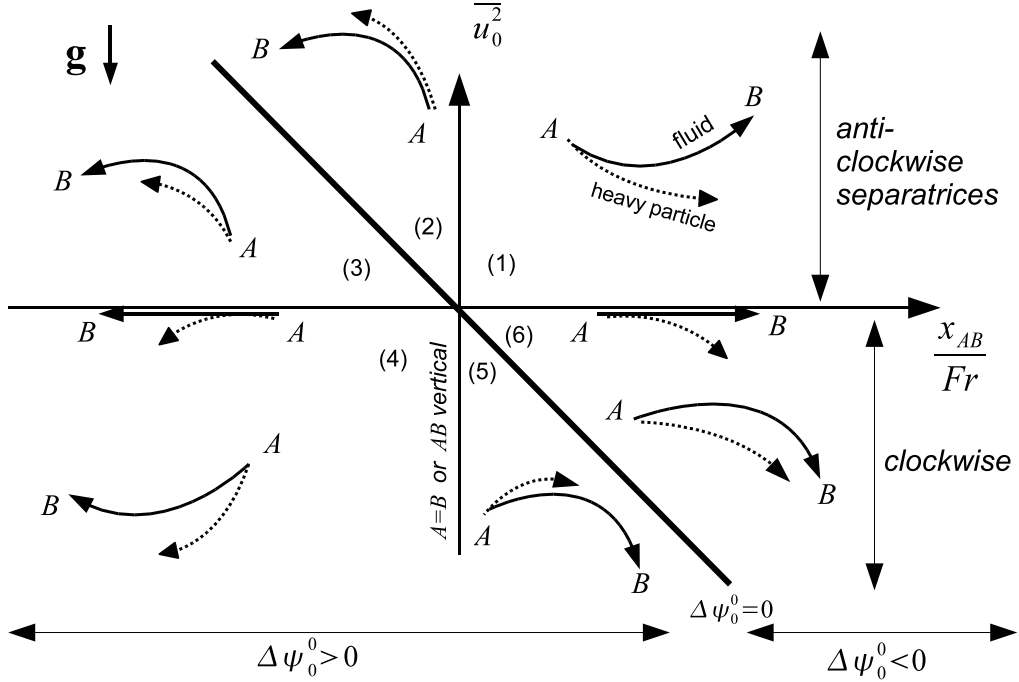}}
\caption{Behavior of particles in the absence of noise, obtained by plotting the sign of $\Delta \psi_0^{0}$ in the plane ($\overline{u^2_0}, x_{AB} / Fr$). 
Solid arcs represent separatrix $AB$, and typical trajectories of heavy inertial particles are sketched in short-dashed. 
}
\label{DiagramNoNoiseHeavyPtcl}
\end{figure}

When $f_0=0$, the sign of  $\Delta \psi_0^0$  allows to determine in which direction the particle will drift nearby the separatrix, in the absence of noise. If $\Delta \psi_0^0 > 0$, the particle will drift towards $+ \mathbf{n}$ (i.e. the left-hand-side of the separatrix), and if $\Delta \psi_0^0 < 0$ it will drift towards $-\mathbf{n}$ (right-hand-side).
Figure \ref{DiagramNoNoiseHeavyPtcl} shows the sign of this quantity in the plane ($\overline{u^2_0},  x_{AB} / Fr$), and enables to distinguish various zones where typical behaviors occur under the effect of the centrifugal force and weight. 
For the sake of clarity, we have only sketched separatrices without inflection points (curvatures have a constant sign). 
In zones (1) and (4) both forces have the same effect (they drive the particle towards the exterior of the separatrix). In  zones (2), (3), (5) and (6), both forces have antagonistic effects. Along the thick oblique line, they balance each other and the particle follows the arc $AB$ in spite of centrifugation and weight.

We now turn to the case $f_0 > 0$, to study the effect of noise on the various behaviors displayed in Fig. \ref{DiagramNoNoiseHeavyPtcl}. We exploit the streamfunction jump $\Delta \psi_0$ of Eq.\ (\ref{Deltapsi01}) as follows. Assuming that the times $\tau_k$, the durations $\delta\tau_k$ and the force $\xi_{kn}$ are independent, then the sum in the right-hand-side of this equation has a zero average value since the random force is isotropic, i.e. it acts in any direction with equal probability. The average of $\Delta \psi_0$ is therefore equal to $\Delta \psi_0^0$. To calculate the fluctuating part we introduce the new random variable: 
\begin{equation}
Z_N = \frac{1}{S_N} \sum_k \delta s_k \, \xi_{kn} ,
\end{equation}
where $N$ is the total number of terms in the sum, and $S_N^2$ is the sum of the variances of each term:
$
S_N^2 = \sum_k \mbox{var}( \delta s_k \, \xi_{kn}).
$
The displacements $\delta s_k$ of $\mathbf{q}$ along the arc $AB$ being independent of $\xi_{kn}$, we have
$
\mbox{var}( \delta s_k \, \xi_{kn}) = \langle \delta s_k^2 \rangle \, \langle \xi^2 \rangle = \langle \delta s_k^2 \rangle,
$
where $\langle \cdot \rangle$ is the average over realizations.
The generalized central limit theorem allows to write that, if $S_N$ satisfies the  
Lyapunov condition, i.e. if there exists some $\delta > 0$ such that $\Lambda_N \equiv \sum_k \langle |\delta s_k \, \xi_{kn}|^{2+\delta} \rangle / S_N^{2+\delta} \to 0$ as $N \to \infty$, then the probability distribution of $Z_N$ converges towards a centered normal distribution with unit variance. 
Under these hypotheses, Eq.\ (\ref{Deltapsi01}) leads to, since $N \gg 1$,
\begin{equation}
\Delta \psi_0 \simeq \Delta\psi_0^0 + \sigma \, Z
\label{ZZ}
\end{equation} 
where $Z$ is a random variable with centered unit normal distribution and $\sigma$ is the standard deviation of $\Delta \psi_0$:
$\sigma = f_0 \, St \, ( \langle\sum \delta s_k^2 \rangle)^{1/2}$.
Eq.\ (\ref{ZZ}) can be used to predict separatrix crossing of inertial particles as follows. Suppose that $\Delta \psi_0^0 > 0$, i.e. in the absence of noise particles with a small inertia will drift towards the  right-hand-side of $AB$. We define as noise-induced crossing the fact that this effect be reversed by the random force, i.e. $\Delta \psi_0 \le 0$. The probability of this event is
\begin{equation}
P(\Delta \psi_0 \le 0) = P\left(  Z \le - \frac{\Delta \psi_0^0}{\sigma} \right)
= \frac{1}{2} \mbox{erfc}\left( \frac{|\Delta \psi_0^0|}{\sigma \,  \sqrt{2}}\right).
\end{equation}
To exploit this result, we approximate the discrete sum of $\delta s_k^2$ by recalling that $\delta s_k = |\mathbf{u}_0(\tau_k-t_0)| \delta \tau_k$, and noticing that $\langle \delta\tau_k^2 \rangle= \beta\, \delta\tau^2$ with $\beta=1$ for constant durations and $\beta=2$ for exponentially distributed random durations.
Then,
\begin{equation}
\langle\sum_k \delta s_k^2 \rangle  = \beta \, \delta \tau \langle \sum_{k}  u_0(\tau_k-t_0)   \delta s_k \rangle \approx \beta \,\delta \tau \int_A^B u_0(s) ds ,
\label{sommecarres}
\end{equation}
where $u_0$ denotes the modulus of $\mathbf{u}_0$. The integral in Eq.\ (\ref{sommecarres}) is simply ${\cal L}_{AB} \, \langle u_0 \rangle_{AB}$ and can be calculated as soon as the fluid velocity field   is known.
Therefore, the variance of $\Delta \psi_0$ reads
\begin{equation} 
\sigma^2 = f_0^2 \, St^2 \, \beta \, \delta \tau \, {\cal L}_{AB} \, \langle u_0 \rangle_{AB}
 = \frac{\beta}{\alpha} \, D \, {\cal L}_{AB} \, \langle u_0 \rangle_{AB} .  
\label{sigma2}
\end{equation} 

The Lyapunov condition can be proved as follows. Writing $u_{0k}$ in place of ${u}_0(\tau_k-t_0)$,  we have, for all $\delta > 0$:
\begin{equation}
\Lambda_N =  \langle |\xi^{2+\delta} |\rangle \frac{\sum_k \langle u_{0k}^{2+\delta} \rangle   \langle \delta\tau_k^{2+\delta} \rangle}{\left( \sum_k \langle u_{0k}^{2} \rangle   \langle \delta\tau_k^{2} \rangle\right)^{1+\delta/2} }.
\label{LambdaN}
\end{equation}
The moment $\langle \delta\tau_k^{2+\delta} \rangle$ in this last equation is equal to $\gamma\, \delta \tau^{2+\delta}$, with $\gamma=1$ for constant durations and $\gamma= \Gamma(3+\delta)$ (where $\Gamma$ denotes Euler's function) for exponentially distributed random durations. 
Let $u_m$ and $u_M$ denote the minimum and maximum moduli of velocities along $AB$ between times $\tau_a$ and $\tau_b$. Then $\Lambda_N$ satisfies:
\begin{equation}
\Lambda_N \le  \langle |\xi^{2+\delta} | \rangle  \, \frac{\gamma}{\beta^{1+\delta/2}}\,  \frac{1}{N^{\delta/2}}
 \left(\frac{u_M} {u_m} \right)^{2+\delta}.
\label{LambdaN2}
\end{equation}
Since the endpoints $\mathbf q(\tau_a)$ and $\mathbf q(\tau_b)$ are fixed and differ from $A$ and $B$, then $u_m$ is finite. Also, the ratio $u_M/u_m$ only depends on the fluid flow and is independent of the number of steps $N$. 
In addition, we assume that $\langle |\xi^{2+\delta} |\rangle$ is finite. 
We then have  
$\Lambda_N \to 0$ as $N \to \infty$, the Lyapunov condition is satisfied and the streamfunction jump $\Delta \psi_0$ has a Gaussian distribution whatever the distribution of the $\boldsymbol \xi$'s.
We conclude that the probability of noise-induced crossing is
\begin{equation}
P(\Delta \psi_0 \le 0)  
= \frac{1}{2} \mbox{erfc}\left( \frac{|\overline{u_0^2}+x_{AB}/Fr|}{ f_0 \, (2  \beta \, \delta \tau \, {\cal L}_{AB} \, \langle u_0 \rangle_{AB} )^{1/2} }\right),
\label{probacrossing}
\end{equation}
in terms of the random force intensity $f_0$, and
\begin{equation}
P(\Delta \psi_0 \le 0)  
= \frac{1}{2} \mbox{erfc}\left( \sqrt{\frac{\alpha}{2\beta}}\frac{St |\overline{u_0^2}+x_{AB}/Fr|}{   ( D \, {\cal L}_{AB} \, \langle u_0 \rangle_{AB} )^{1/2} }\right),
\label{probacrossingD}
\end{equation}
in terms of the diffusion coefficient $D$. This expression for the probability of noise-induced crossing is the main result of this work, and
will be compared to numerical solutions in the next paragraphs.


\section{Application to a circular dividing streamline}

\begin{figure}[h] 
{\includegraphics[scale=0.7]{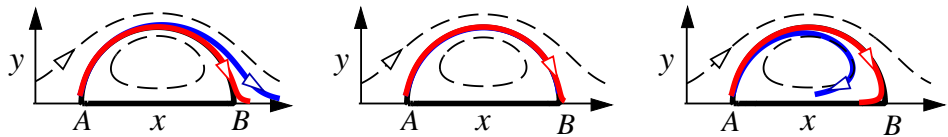}} 
\caption{Deterministic particle trajectories in a test flow corresponding to three zones of the diagram of Fig. \ref{DiagramNoNoiseHeavyPtcl}, obtained by solving Eq.\
(\ref{ptclDynAdim}) numerically. Left: centrifugal effects dominate (zone (5) of Fig. \ref{DiagramNoNoiseHeavyPtcl}). Middle: centrifugation and gravity balance each other (oblique line of Fig. \ref{DiagramNoNoiseHeavyPtcl}). Right: gravity dominates centrifugation (zone (6)). Two test particles heavier than the fluid have been used, with two different densities.}
\label{TroisTrajsSciLabXFIG}
\end{figure}

To illustrate these results, we have  conducted a series of numerical simulations by solving Eq.\ (\ref{ptclDynAdim}) with the following elementary flow:
$
\psi_0(x,y) = 2  \, y \, (x^2+y^2 - 1/4).
$
It corresponds to a circular cell (Fig.\ \ref{TroisTrajsSciLabXFIG}) with unit diameter and peak velocity equal to 1 on the dividing streamline at $(x,y)=(0,1/2)$. 
We  focus on the upper separatrix of this cell, and get: $\overline{u_0^2} = -{\pi} / 2$ and $\langle u_0 \rangle_{AB} = 2/\pi$. Here, $\overline{u_0^2} < 0$, in agreement with the fact that this separatrix is clockwise. Therefore, since $x_{AB} > 0$ in this case, the flow is in the lower-right quadrant of Fig. \ref{DiagramNoNoiseHeavyPtcl}.
We consider three typical cases, corresponding to three zones of this diagram:
  $Fr > \sqrt{2/\pi}$ (i.e. $\Delta\psi_0^0 > 0$, zone (5)),  $Fr = \sqrt{2/\pi}$ (oblique line $\Delta\psi_0^0 = 0$), and
$Fr < \sqrt{2/\pi}$ (zone (6)). 
The three cases are plotted in Fig.\ \ref{TroisTrajsSciLabXFIG},
 for $Fr = 1.43$ (left graph), $Fr = \sqrt{2/\pi} \approx 0.637 $ (center graph),  $Fr = 0.159$ (right graph). 
Two test particles have been used, corresponding to a density ratio $\gamma_p = 6500$ (blue line) and 
 $\gamma_p = 417$ (red line).  
 These values correspond, for example, to iron and wood particles of radius $a=5 \, \mu m$, driven by an air flow near a recirculation cell of diameter $5 \, cm$, at velocities less than one $m.s^{-1}$.  
 In the six runs, the corresponding Stokes numbers $St$ of both types of particles are: $St=0.040$ and $0.0026$ (left graph of Fig.\ \ref{TroisTrajsSciLabXFIG}), $St=0.027$ and $0.0017$ (center), $St=0.013$ and $0.0009$ (right). 
 We observe that particles behave as predicted by the diagram of Fig. \ref{DiagramNoNoiseHeavyPtcl}. 

\begin{figure}[h] 
\hspace*{-.5cm}{\includegraphics[scale=0.6]{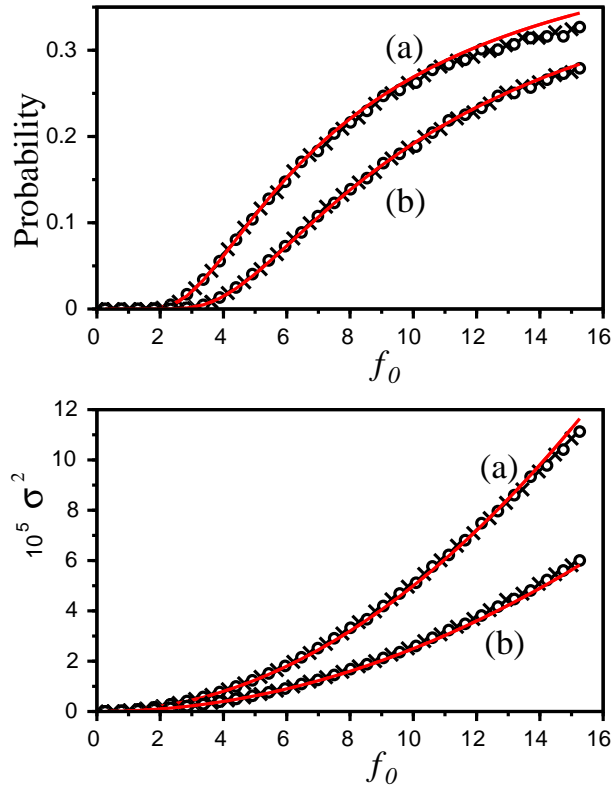}} 
\caption{Upper graph: 
 theoretical probabilities (red lines) given by Eq.\ (\ref{probacrossing}). (a): random $\delta \tau_k$ ; (b): constant $\delta \tau_k$.
Symbols are the probabilities obtained from numerical simulations. Circles: Gaussian force intensity $\xi_{ik}$ ; crosses: uniform $\xi_{ik}$.    Lower graph: variance of the streamfunction jump, same symbols. Red solid lines are the theoretical law (\ref{sigma2}).}
\label{Probaf0Sigma2f0XFIG}
\end{figure}

Results of numerical simulations in the non-deterministic case ($f_0 > 0$) are shown in Fig.\ \ref{Probaf0Sigma2f0XFIG}. Here, $5. \, 10^4$ 
particles have been injected at the same point, slightly above $A$, and tracked until they reach the vicinity of $B$. 
The Froude number is $Fr=1.43$, and corresponds to zone (5) of Fig.\ \ref{DiagramNoNoiseHeavyPtcl}, i.e. centrifugal effects dominate gravity. The Stokes number is  $St=0.005$ in all cases.
Then, according to the sign of $\Delta \psi_0$ for each particle, the probability of noise-induced crossing has been computed. 
Two kinds of distributions have been used for $\xi_i$:
normalized Gaussian distribution (circles in Fig.\ \ref{Probaf0Sigma2f0XFIG}) and uniform distribution between $-\sqrt 3$ and $\sqrt 3$ (crosses). Both have a unit variance. The intensity of the random force $f_0$ has been varied  from 0 to 15. 
Also, the time scale of noise is taken to be $\delta\tau = 0.01$. Two kinds of durations are used: exponentially distributed durations with mean $\delta\tau$ (curves (a)), and constant durations $\delta\tau_k=\delta\tau$ (curves (b)).
We have checked that the distribution of $\Delta\psi_0$ is very close to Gaussian, even if $\boldsymbol \xi$ is not, in agreement with the central limit theorem.
We observe that numerical simulations agree with the theoretical result of Eq.\ (\ref{probacrossing}), for both kinds of forcing (uniform or Gaussian). Because this agreement relies on the validity of approximation (\ref{sigma2}), we have also plotted the numerical value of $\sigma^2$ versus $f_0$. The agreement is correct, at least for small or moderate $f_0$'s, as expected. 

\section{Conclusion}

The theoretical analysis presented here allows to predict separatrix crossing due to an additive random force, for particles transported in a two-dimensional steady flow, without solving any particle trajectory. It can be generalized to multiplicative noise by including the (space-dependent) force intensity $f_0$ into the averaged terms. This might be useful for the application of this work to turbulent separatrix crossing modeled by means of the eddy lifetime model.
Also, the theory can be used not only for analytical flows, but also for more complex fluid flows, like those obtained in computational fluid dynamics. Indeed, as soon as the mean flow is known, the kinematic quantities $\overline {u_0^2}$ and $\langle u_0 \rangle_{AB}$ can be easily computed and included into formula (\ref{probacrossing}). This was done in Ref.\ \cite{Verjus2016} for deterministic time-periodic perturbations, and can be done also for the stochastic case considered here. The applicability of this analysis to experimental flows, using Particle Image Velocimetry measurements to accurately describe the steady component ${\mathbf u}_0$, is a promising perspective that will constitute the next step of this study. 

Result (\ref{probacrossing}) can be readily extended to the case where the particle-to-fluid density ratio is no longer large, provided the Basset term can be neglected in the Maxey-Riley equations \cite{Maxey1983}. This would be especially useful for solid particles transported in liquids, either heavier or lighter than the fluid.
The diagram of Fig.\ \ref{DiagramNoNoiseHeavyPtcl} can be easily generalized to the case where buoyancy and pressure gradient force of the undisturbed flow (Tchen's force) are present. 
  However, it is well known that, for such particles, forces due to the unsteadiness (Basset's force) and to the inertia (lift force \cite{Saffman1965}) of the disturbed flow can affect separatrix crossing and should be introduced rigorously (see also the work by Druzhinin and Ostrovsky \cite{Druzhinin1994} for the effect of Basset's force on separatrix crossing). No general definite expressions for these forces are available so far, and this is the reason why we have chosen to limit ourselves to the sole case of aerosols. The generalization of the present analysis to particles of density comparable to that of the fluid is among the perspectives of this work.

%

\end{document}